\documentclass[twoside,fleqn]{article}
\usepackage{npb,axodraw2,amsmath,moreverb}

\allowdisplaybreaks

\renewcommand{\listingoffset}{1em}

\newcommand{\eg}{e.g.\ }
\newcommand{\ie}{i.e.\ }

\newcommand{\ord}{\mathcal{O}}
\newcommand{\M}{\mathcal{M}}
\newcommand{\ri}{\mathrm{i}}

\newcommand{\lbrac}{\symbol{123}}
\newcommand{\rbrac}{\symbol{125}}
\newcommand{\uscore}{\symbol{95}}

\newcommand{\ind}{\hspace*{\parindent}}

\newcommand{\ket}[1]{\left| #1\right\rangle}
\newcommand{\bra}[1]{\left\langle #1\right|}
\newcommand{\braket}[2]{\left\langle #1\vphantom{#2}%
  \right. \kern-2.5pt\left| #2\vphantom{#1}\right\rangle}

\newcommand{\Tr}{\mathop{\mathrm{Tr}}}
\newcommand{\unity}{\mathrm{1\mskip-4.25mu l}}
\newcommand{\diracslash}[1]{\setbox0=\hbox{$#1$}%
  \rlap{\ifdim\wd0>.7em\kern.22\wd0\else\kern.1\wd0\fi /}#1}

\newcommand{\sigbar}{\overline{\sigma}}
\makeatletter
\newcommand\teenyweeny{\@setfontsize\teenyweeny{6}{5}}
\makeatother
\newcommand{\sigpbar}{%
  \rlap{\kern -.175em\raise 1.4ex\hbox{\teenyweeny$_{(\;\,\,)}$}}%
  \sigbar}

\newcommand{\lhsFierz}[2]{%
  \unitlength=1bp%
  \vcenter{\hsize=60\unitlength%
  \begin{picture}(50,53)
  \SetScale{.5}
  \SetOffset(-15,0)
  \SetWidth{1}
  \Line(25,85)(115,85)
  \Text(20,85)[r]{$A$}
  \Text(120,85)[l]{$B$}
  \Line(25,25)(115,25)
  \Text(20,25)[r]{$C$}
  \Text(120,25)[l]{$D$}
  \Photon(70,25)(70,85){2}{5}
  \Vertex(70,25){3}
  \Vertex(70,85){3}
  \Text(72,93)[b]{$#1$}
  \Text(72,17)[t]{$#2$}
  \end{picture}}
}
\newcommand{\rhsFierzDelta}{%
  \unitlength=1bp%
  \vcenter{\hsize=60\unitlength%
  \begin{picture}(50,53)
  \SetScale{.5}
  \SetOffset(-15,0)
  \SetWidth{1}
  \Line(25,85)(57,85)
  \Line(57,85)(83,25)
  \Line(83,25)(115,25)
  \Text(20,85)[r]{$A$}
  \Text(120,25)[l]{$D$}
  \Line(25,25)(57,25)
  \Line(57,25)(83,85)
  \Line(83,85)(115,85)
  \Text(120,85)[l]{$B$}
  \Text(20,25)[r]{$C$}
  \end{picture}}
}
\newcommand{\rhsFierzEps}{%
  \unitlength=1bp%
  \vcenter{\hsize=60\unitlength\hfill%
  \begin{picture}(50,53)
  \SetScale{.5}
  \SetOffset(-15,0)
  \SetWidth{1}
  \Line(25,85)(57,85)
  \Line(57,85)(57,25)
  \Line(57,25)(25,25)
  \BBox(50,48)(64,62)
  \Text(57,55)[]{$\varepsilon$}
  \Text(20,85)[r]{$A$}
  \Text(20,25)[r]{$C$}
  \Line(115,85)(83,85)
  \Line(83,85)(83,25)
  \Line(83,25)(115,25)
  \BBox(76,48)(90,62)
  \Text(83,55)[]{$\varepsilon$}
  \Text(120,85)[l]{$B$}
  \Text(120,25)[l]{$D$}
  \end{picture}}
}

\newenvironment{smatrix}%
  {\left(\begin{matrix}}%
  {\end{matrix}\right)}

\begin{document}

\title{Optimizations for the Computation of Radiative Corrections}

\author{Thomas Hahn\address{%
	Max-Planck-Institut f\"ur Physik\hfill {\small MPI-PhT/2002-46} \\
        F\"ohringer Ring 6 \\
	D--80805 Munich, Germany}}

\begin{abstract}
Two methods are presented with which the CPU time spent on the calculation
of radiative corrections can be significantly reduced.  The first is the
parallelization of the program, which can be surprisingly simple to
implement under certain circumstances often met in the calculation of
radiative corrections.  The second is the efficient direct calculation of
fermion chains.  The latter not only improves the overall performance of
the program, but introduces better conceptual clarity as well, as it
allows for a homogeneous treatment of bosonic and fermionic amplitudes.
\end{abstract}

\maketitle


\section{Introduction}

The calculation of radiative corrections is in general a very CPU-time
consuming business, especially if ``aggravating conditions'' are met, such
as the scans over large areas in parameter space typical for extensions of
the Standard Model.

The running time of a program that computes radiative corrections can be 
written as
\begin{equation}
T = N\tau + \varepsilon\,,
\end{equation}
where $N$ is the number of points in phase and parameter space and $\tau$
is the time to compute one such point.  The remainder $\varepsilon$
accounts for all code which is executed only $n\ll N$ times, like
initializations of model parameters, etc., and is thus generally
negligible.  Typical numbers are $N\sim 10^5\dots 10^7$ and
$\tau\sim 100$~ms.

The two techniques presented in this contribution reduce CPU time via
both factors:
\begin{itemize}
\item
Parallelization of the program reduces execution time from $\ord(N)$ to 
$\ord(N/N_{\text{processors}})$,

\item
Directly computing fermion chains (in processes involving external 
fermions), rather than using the standard trace technique, reduces $\tau$ 
from $\ord(n_F^2)$ to $\ord(n_F)$, where $n_F$ is the number of fermionic 
structures.
\end{itemize}

The following discussion is limited to numerical programs.  The examples
are written in Fortran~77, which is not only still in wide use among
numerical programmers, but moreover constitutes a ``lowest common
denominator'' in the sense that if a technique can be implemented in
Fortran~77, it can be implemented in almost any language.


\section{Parallelization}

Parallelization in general is a very difficult topic, and considerable
effort has been spent on the development of sophisticated libraries, 
compilers, and other tools.  In the business of computing radiative
corrections, however, programs often contain ``naturally'' parallel
structures, for example if the same calculation is performed for different
values of a parameter, viz.
\begin{verbatim}
  do tanBeta = 2, 20
    xsec = CrossSection(tanBeta)
    print *, tanBeta, xsec
  enddo
\end{verbatim}
Such programs are known as \emph{essentially parallel}, as each cycle of
the loop may be executed independently without having to rearrange any
code.  In contrast, a parallelizing compiler will generally have to apply
advanced techniques such as loop interchange and cache tiling before
arriving at parallelizable code (see \eg\cite{pgi}).

The parallelization of such loops turns out to be surprisingly simple on a
symmetric multiprocessing (SMP) architecture, that is, a single machine
with several processors.  Driven by demand in areas such as commercial Web
hosting, SMP machines have meanwhile become affordable even for
cash-strapped institutes and are well supported by all major operating
systems.

The two basic operations are \verb=fork= and \verb=wait=:  A \verb=fork=
creates an exact copy of the process that calls it; it returns 0 to the
child process and the child's process id to the parent.  The opposite is
\verb=wait=, which suspends the caller until one of its child processes
has terminated.

Although \verb=fork= and \verb=wait= are not strictly part of the ANSI
Fortran-77 Standard, they are available on many Unix systems.\footnote{%
One notable exception is the g77 compiler.  It is not difficult to work 
around this, however: set up a short C program, wrapper.c, which 
contains\\
\begin{tt}
\ind \#include <unistd.h> \\
\ind \#include <sys/wait.h> \\
\ind int fork\uscore() \lbrac\ return fork(); \rbrac \\
\ind int wait\uscore(int *status) \lbrac\ return wait(status); \rbrac \\
\end{tt}
and add wrapper.c to the g77 command line, viz.\\
\ind g77 (options) myprogram.f wrapper.c (libs)
}  They belong to a library of wrapper functions that 
provide Fortran access to certain functions of the C library.  These 
functions are sometimes known also as the ``3F'' functions because they 
are described in section 3F of the Unix man pages.

The implementation is straightforward: The parent process executes the
main loop and in each cycle of the loop forks off a child process to do
the actual calculation.  The concurrent processes are automatically sent
on the available processors by the operating system.

The reason why the implementation is so simple is that the forked
processes run \emph{completely independently}.  This means that except for
screen (and possibly file) output there are no resource-sharing issues to
take care of, like avoiding simultaneous write access to some variable.  
By the same token, there is no simple way (in pure Fortran\footnote{Two
possibilities are: 1)~Regular Fortran file I/O, but with named pipes.  
2)~Data exchange via Unix domain sockets, although this requires some
minor mixed-lan\-guage programming in C.}) for the child process to
communicate its results back to the parent.  If the results only have to
be stored in a file, however, such problems can simply be solved by
opening a unique output file for each child process and redirecting
standard output to that file if necessary.

Consider the following example program:
\begin{listing}{1}
seqnum = 0
processes = 0
cpus = 4

do tanBeta = 2, 20
  if( processes .lt. cpus ) then
    processes = processes + 1
  else
    pid = wait(0)
  endif
  seqnum = seqnum + 1

  if( fork() .eq. 0 ) then
    write(log, '("log.",I5.5)') seqnum
    open(6, file=log)
    xsec = CrossSection(tanBeta)
    print *, tanBeta, xsec
    return
  endif
enddo

do i = 1, processes
  pid = wait(0)
enddo
\end{listing}

The main do loop ($\ell$.~5--20) has been extended with respect to the
serial version and falls into two blocks.  The first block ($\ell$.~6--11)
is executed by the parent only.  The parent calls \verb=fork= in
$\ell$.~13 and immediately skips to the next cycle of the loop.  The
statements inside the \verb=if= block ($\ell$.~14--18) are executed by the 
child process only.

The parent takes care not to spawn more child processes than there are
processors available, so as not to overload the system.  The number of
concurrently running child processes is stored in the variable
\verb=processes= and the number of processors in \verb=cpus=.  The first
\verb=cpus= cycles of the loop ``fill up'' the available processors with
jobs.  After that the parent waits for a running child process to finish
before proceeding to the next \verb=fork= ($\ell$.~9).

The child process, for which \verb=fork= returns 0, branches into the
\verb=if= block.  It first generates a unique file name using a sequence
number ($\ell$.~14) which it then opens as unit 6, thus redirecting
terminal output to this file.  In this way it is guaranteed that the
output of concurrent processes is not mixed up.  Most time is spent on the
actual calculation ($\ell$.~16--17) which is identical to the serial
version.  Unlike in the serial version, however, the child process does
not return to the main loop, but terminates at the \verb=return=
statement ($\ell$.~18).

After completing the main loop, the parent has to wait until all child
processes have finished ($\ell$.~22--24).

The overhead incurred by a \verb=fork= is negligible unless very short
units are parallelized.  \verb=fork= is a very common operation on
preemptive multitasking systems such as Unix and is therefore usually
implemented very efficiently.  The Linux kernel, for instance, uses
copy-on-write memory pages, that is, a memory page is not duplicated until
either parent or child write on it.  In fact, it will be difficult to
notice any performance penalty of the parallel version running on a single
CPU compared to the serial version.

If care is taken not to overload the system with more child processes than
there are processors available (which would only require unnecessary task
switches and use up memory resources) it is reasonable to expect a
near-optimal speed-up, \ie $T\to T/N$ on $N$ processors.


\section{Fermionic Objects}

Amplitudes of processes with external fermions can be written as
\begin{equation}
\M = \sum_{i = 1}^{n_F} c_i\, F_i
\end{equation}
where the $F_i$ are (products of) fermion chains, \ie are of the form
$\prod \bra{u}\Gamma_i\ket{v}$, where $u$ and $v$ are spinors, and
$\Gamma_i$ is a product of Dirac matrices.

One usually proceeds to compute probabilities, \eg $|\M|^2$, rather than 
the amplitude $\M$ itself, because one can then use the trace technique:
\begin{equation}
|\M|^2 = \sum_{i, j = 1}^{n_F} c_i^*\, c_j\, F_i^* F_j
\end{equation}
where $F_i^* F_j$ is computed as
\begin{equation}
\begin{split}
F_i^* F_j
&= \bra{v}\bar\Gamma_i\ket{u} \bra{u}\Gamma_j\ket{v} \\
&= \Tr\bigl(\bar\Gamma_i\,\ket{u}\!\bra{u}\,
            \Gamma_j\,\ket{v}\!\bra{v}\bigr)\,.
\end{split}
\end{equation}
The advantage is that no explicit representation of the spinors is needed 
since the projection operators can be expressed through Dirac matrices 
only, \eg $\ket{u_\lambda}\bra{u_\lambda} = \frac 12 (\unity +
\lambda\gamma_5) \diracslash{p}$ for a massless fermion.

The problem with the trace technique is that it scales as $n_F^2$.  That
is, one needs to compute all $n_F^2$ combinations of $F_i^* F_j$ and not
just $n_F$ of the $F_i$ as for $\M$ alone.  For $n_F = 20$, for example,
one has to compute 400 traces.  More severely, matters get worse the more
vectors are in the game, \eg in multi-particle final states, or with
polarization effects, because generally all combinations of vectors can
appear in the $\Gamma_i$, and thus $n_F\sim (\text{number of vectors})!$.

The obvious alternative to the trace technique is to insert the
4-dimensional representation of the spinors and the Dirac matrices and
work out the matrix algebra.  This procedure is not entirely
straightforward to implement in a language like Fortran, however, because
in general Lorentz indices connect different fermion chains.  Also, the
calculational efficiency is not quite optimal because the 4-dim.\
representation contains redundancy, \eg entire $2\times 2$ blocks in the
Dirac matrices are zero.

Actually, the 4-dim.\ representation is used for mathematical convenience
only and, from the physical point of view, fermions are indeed more
naturally represented by 2-dim.\ objects which come in two kinds, left-
and right-handed.  This makes the Weyl representation an obvious choice.

The Weyl representation can in principle be used from the very beginning,
\ie already at the level of the Feynman rules \cite{dittmaier98}.  This is
only somewhat problematic for loop calculations, where regularization is
needed, since it is not obvious how to extend the 2-dim.\ objects
appearing in the Feynman rules to $D$ dimensions.  In the present approach
therefore everything is kept 4 (or $D$) dimensional during the algebraic
simplification, such that \eg $\gamma_\mu\gamma^\mu\to D$ can be replaced,
and the Weyl representation is inserted only at the very end, just before
the numerical evaluation, and only for objects involving external fermions
(\ie not for internal fermion traces).

The Dirac matrices are given in the Weyl representation by
\begin{equation}
\begin{aligned}
\gamma_\mu = \begin{pmatrix}
	0 & \sigma_\mu \\
	\sigbar_\mu & 0
	\end{pmatrix}, \quad
\omega_+ &= \begin{pmatrix}
	\unity & 0 \\ 
	0 & 0
	\end{pmatrix}, \\
\omega_- &= \begin{pmatrix}
	0 & 0 \\
	0 & \unity
	\end{pmatrix},
\end{aligned}
\end{equation}
where the $\sigpbar_\mu$ matrices are defined in terms of the ordinary 
Pauli matrices $\vec\sigma = (\sigma_1, \sigma_2, \sigma_3)$ as
\begin{equation}
\sigma_\mu = (\unity, -\vec\sigma)\,, \quad
\sigbar_\mu = (\unity, +\vec\sigma)\,.
\end{equation}
Introducing furthermore 2-dim.\ spinors via
\begin{equation}
\bra{u} = \bigl(\bra{u_+}, \bra{u_-}\bigr)\,,
\quad
\ket{v} = \begin{pmatrix}
\ket{v_-} \\
\ket{v_+} 
\end{pmatrix},
\end{equation}
it is a simple exercise to show that every \emph{chiral} 4-dim.\ Dirac 
chain can be converted to a \emph{single} 2-dim.\ sigma chain:
\begin{align}
\bra{u}\omega_-\gamma_\mu\gamma_\nu\cdots\ket{v}
&= \bra{u_-}\sigbar_\mu\sigma_\nu\cdots\ket{v_\pm}, \\
\bra{u}\omega_+\gamma_\mu\gamma_\nu\cdots\ket{v}
&= \bra{u_+}\sigma_\mu\sigbar_\nu\cdots\ket{v_\mp}.
\end{align}
That is, going from the 4-dim.\ to the 2-dim.\ representation does not 
increase $n_F$.

The two most important identities satisfied by the $\sigpbar$ matrices are 
the Fierz identities \cite{core}
\begin{gather}
\notag
\bra{A}\sigma_\mu\ket{B} \bra{C}\sigbar^\mu\ket{D}
= 2\braket{A}{D} \braket{C}{B}, \\[1ex]
\lhsFierz{\sigma^\mu}{\sigbar_\mu} \;\;=\;\; 2\;\rhsFierzDelta \\[3ex]
\notag
\kern -5pt
\bra{A}\sigpbar_\mu\ket{B} \bra{C}\sigpbar^{\,\mu}\ket{D}
= 2\bra{A}\varepsilon\ket{C} \bra{B}\varepsilon\ket{D},
\\[1ex]
\lhsFierz{\sigpbar^{\,\mu}}{\sigpbar_\mu} \;\;=\;\; 2\;\rhsFierzEps
\end{gather}
where $\varepsilon = \left(\begin{smallmatrix} 0 & 1 \\ -1 & 0 
\end{smallmatrix}\right)$ is the spinor metric.

These identities allow to completely remove all Lorentz contractions
between sigma chains.  For the implementation in a computer program this
is a decisive simplification, for it means that after application of the
Fierz identities one can calculate one sigma chain at a time, independent
of any other sigma chains.

To make the implementation even simpler, one can express also the
remaining kinematical objects through $\sigpbar$ according to
\begin{gather}
g_{\mu\nu} = \frac 12\Tr(\sigma_\mu\sigbar_\nu)\,, \\
\notag
\varepsilon_{\lambda\mu\nu\rho}
= \frac 12\Tr(\sigma_\lambda\sigbar_\mu\sigma_\nu\sigbar_\rho)
- g_{\lambda\mu} g_{\nu\rho}
\\ \qquad\qquad{}
+ g_{\lambda\nu} g_{\mu\rho}
- g_{\lambda\rho} g_{\mu\nu}\,.
\end{gather}
Now the four-vectors themselves are no longer needed, but only their 
contractions with $\sigpbar$,
\begin{gather}
\sigma_\mu k^\mu = \begin{pmatrix}
  k^0 + k^3     & k^1 - \ri k^2 \\
  k^1 + \ri k^2 & k^0 - k^3
\end{pmatrix} =: \hat k\,, \\
\sigbar_\mu k^\mu = \begin{pmatrix}
  k^0 - k^3      & -k^1 + \ri k^2 \\
  -k^1 - \ri k^2 & k^0 + k^3
\end{pmatrix} =: \hat{\overline{k}}\,.
\end{gather}

Altogether now, the following objects and operations need to be 
implemented:
\begin{itemize}
\item arrays: all spinors, all vectors:
\begin{align}
\ket{u_\pm} &\sim \begin{smatrix}
  u_1\vphantom{b} \\
  u_2\vphantom{d}
\end{smatrix}, \\
\hat k,\ \hat{\overline{k}} &\sim \begin{smatrix}
  a & b \\
  c & d
\end{smatrix}.
\end{align}

\item functions: spinor $\times$ spinor (\texttt{ss}), matrix $\times$ 
spinor (\texttt{ms}), matrix $\times$ matrix (\texttt{mm}):
\begin{align}
\braket{u}{v} &\sim
\left(u_1\hskip\arraycolsep u_2\right)\cdot\begin{smatrix}
  v_1\vphantom{b} \\
  v_2\vphantom{d}
\end{smatrix}, \\
\hat k\ket{v} &\sim \begin{smatrix}
  a & b \\
  c & d
\end{smatrix}\cdot\begin{smatrix}
  v_1\vphantom{b} \\
  v_2\vphantom{d}
\end{smatrix}, \\
\hat k_1\hat k_2 &\sim \begin{smatrix}
  a_1 & b_1 \\
  c_1 & d_1
\end{smatrix}\cdot\begin{smatrix}
  a_2 & b_2 \\
  c_2 & d_2
\end{smatrix}.
\end{align}
These almost trivial functions are sufficient to compute arbitrary sigma 
chains by repeated application, \eg
\begin{gather}
\bra{u}\hat k_1\hat{\overline{k}}_2\hat k_3\ket{v} =
\\ \notag
\mathtt{ss}( u,\,
  \mathtt{ms}(\hat k_1,\,
    \mathtt{ms}(\hat{\overline{k}}_2,\,
      \mathtt{ms}(\hat k_3,\, v
))))\,.
\end{gather}
Note that in order to nest the functions inside each other as above, the
matrix-valued functions have to write their results to an intermediate
array (a kind of accumulator register) and return the location of that
array.
\end{itemize}

Finally, one should note that when the fermion chains are computed
directly, the amplitude $\M$ becomes a handy complex number, rather than a
``bunch of form factors,'' $\{c_i F_i\}$.  That is, to compute for example
$|\M|^2$, one does not need to sum up a list of numbers, but only take the
square of a complex variable.  In this sense fermionic amplitudes are now
treated on the same footing as bosonic amplitudes.  Furthermore, since the
$c_i$ no longer have to be factored out in front of the $F_i$, the
analytic expression for $\M$ may be rearranged freely to yield the most
compact form.


\section{Summary}

Two methods of speeding up the calculation of radiative corrections have 
been demonstrated.

\paragraph{Parallelization}

Programs for computing radiative corrections are often natural candidates
for parallelization.  It is not difficult to attain near-optimal speed-up,
\ie $T\to T/N$ on $N$ processors.

On SMP machines, it is very easy to parallelize Fortran programs with only
few changes in the code.  The two basic functions \verb=fork= and
\verb=wait= are available by default in many Fortran libraries, and even
if not, only a trivial wrapper program in C is needed.

\paragraph{Direct calculation of fermion chains}

The number of fermionic structures that have to be computed can be reduced
from $\ord(n_F^2)$, with the trace technique, to $\ord(n_F)$ when
calculating the fermion chains directly.  Moreover, with the latter
approach one gets a simple complex number for the amplitude $\M$.

Using the Weyl representation, the computation of the fermion chains is
broken down from a 4-dim.\ to a 2-dim.\ problem.  The Fierz identities
allow to completely disentangle fermion chains connected by Lorentz
indices.

The implementation is simple, even more so if one systematically expresses
all kinematic objects through sigma chains, also non-fermionic ones like
$g_{\mu\nu}$, $\varepsilon_{\lambda\mu\nu\rho}$, $k_\mu$.

\end{document}